\documentclass[a4paper,12pt]{article} 
\pdfoutput=1
\usepackage{pdfpages}
\usepackage{graphicx}
\usepackage{amsmath, amsthm, amssymb}
\usepackage{setspace}  
\usepackage[super]{natbib}
\bibpunct{}{}{,}{a}{,}{,}

\usepackage{epsfig}

\def\thesection{\Roman{section}}

\newcommand{\be}{\begin{equation}}
\newcommand{\ee}{\end{equation}} 
\newcommand{\lb}{\label}
\newcommand{\ol}{\overline}

\newcommand{\btau}{{\mbox{\boldmath $\tau$}}}

\newcommand{\bk}{{\bf k}}

\newcommand{\bp}{{\bf p}}
\newcommand{\br}{{\bf r}}
\newcommand{\bu}{{\bf u}}
\newcommand{\bw}{{\bf w}}

\newcommand{\bv}{{\bf v}}
\newcommand{\bx}{{\bf x}}

\newcommand{\fltr}{\overline}
\newcommand{\wt}{\widetilde}

\newcommand{\fu}{\fltr\bu}

\newcommand{\grad}{{\mbox{\boldmath $\nabla$}}}
\newcommand{\bdot}{{\mbox{\boldmath $\cdot$}}}
\newcommand{\bdots}{{\mbox{\boldmath $:$}}}

\newcommand{\bzed}{{\mbox{\boldmath $0$}}}

\begin{document}             




\baselineskip=18pt
\renewcommand{\thefootnote}{\alph{footnote})}

\begin{center}
\begin{Large}
{\bf Localness of energy cascade in hydrodynamic turbulence, I.
Smooth coarse-graining} \\
\end{Large}

\bigskip
\bigskip

Gregory L. Eyink\footnote{Email: eyink@ams.jhu.edu}
and Hussein Aluie\footnote{Email: hussein@jhu.edu} \\
{\it
Department of Applied Mathematics \& Statistics\\ 
The Johns Hopkins University \\
3400 North Charles Street \\
Baltimore, MD 21218-2682\\
}

\bigskip
\bigskip

\begin{abstract}

We introduce a novel approach to scale-decomposition of the fluid kinetic energy 
(or other quadratic integrals) into band-pass contributions from a series of 
length-scales. Our decomposition is based on a multiscale generalization of the 
``Germano identity'' for smooth, graded filter kernels. We employ this method 
to derive a budget equation that describes the transfers of turbulent  kinetic
energy both in space and in scale. It is shown that the inter-scale energy 
transfer is dominated by local triadic interactions, assuming only the scaling 
properties expected in a turbulent inertial-range.  We derive rigorous upper 
bounds on the contributions of non-local triads, extending the work of Eyink
(2005)\cite{Eyink05} for low-pass filtering. We also propose a physical explanation 
of the differing exponents for our rigorous upper bounds and for the scaling predictions
of Kraichnan (1966,1971)\cite{Kraichnan66,Kraichnan71}. The faster decay
predicted by Kraichnan is argued to be the consequence of additional
cancellations in the signed contributions to transfer from non-local triads,  
after averaging over space. This picture is supported by data from a 
$512^3$ pseudospectral simulation of Navier-Stokes turbulence with
phase-shift dealiasing. 

\end{abstract}
\end{center}

\vspace{0.5cm}


\clearpage

\clearpage

\newpage


\section{Introduction}

The traditional Richardson-Kolmogorov-Onsager picture of turbulence supposes an 
energy cascade local in scale. That is, the triad interactions responsible for energy flux 
across a given length-scale are supposed to involve three modes all of comparable scale, 
differing at most by some fixed ratio. This statement also implies that the energy transfer  is 
between modes at roughly the same scale.  The scale-locality property is fundamental to justify 
the concept of  universality of small-scale statistics in turbulent flow.  In a recent work \cite{Eyink05} 
one of us has given  a rigorous  proof of the scale-locality property for solutions of the 
Navier-Stokes equation, based on the sole assumption that the velocity field has the 
regularity/scaling properties observed in turbulent flow. The argument employed a smooth 
coarse-graining (or low-pass filtering) technique to resolve the velocity fields simultaneously 
both in space and in scale.  The results in Eyink (2005)\cite{Eyink05} apply to individual 
Navier-Stokes solutions, with no statistical averaging, pointwise in space.  They therefore 
give substantial support to the classical picture of a local energy cascade.  

\textcolor{black}
{Recent numerical studies 
\cite{Alexakisetal05b,Mininni06, Mininni08,Domaradzki07a,Domaradzki07b,Domaradzki09}
have largely supported these conclusions about energy flux. All of these groups verify the 
individual dominance of nonlocal triadic interactions but conclude that the net energy flux 
from summation over triads is dominated by local interactions.} In particular, the studies
\cite{Domaradzki07a,Domaradzki07b,Domaradzki09} calculated the locality function 
$W(s)$ of Kraichnan \cite{Kraichnan66,Kraichnan71} which measures the contribution 
to energy flux from wavenumber triads $(k,q,p)$ with a fixed scale-ratio $s=\min\{k,q,p\}/
{\rm med}\{k,q,p\}.$ They verify to reasonable accuracy the prediction of Kraichnan that $W(s)
\propto s^{4/3}$ for $s\ll 1,$ in agreement with some earlier DNS results \cite{Zhou93a,Zhou93b}. 
Note that this decay of nonlocal contributions in $s$ is even faster than that proved in Eyink
(2005)\cite{Eyink05},  which gives $W(s)=O(s^{2/3})$ as a rigourous upper bound. The 
studies \cite{Domaradzki07a,Domaradzki07b,Domaradzki09} find also no difference between 
graded and sharp filters, at least with bands defined for a geometric sequences of scales.

Nevertheless, debate continues on this important issue. In addition to using low-pass filtering 
as in Eyink (2005)\cite{Eyink05}, the numerical works \cite{Alexakisetal05b,Mininni06, Mininni08,Domaradzki07a,Domaradzki07b,Domaradzki09} have studied  the problem 
also by using {\it band-pass filtering}. These studies employed both sharp-spectral and 
graded kernels to define modal bands. \textcolor{black}{The relevant nonlinear transfer 
functions for band-pass fields are no longer energy flux but {\it interband energy transfers}.
In this setting, one group \cite{Alexakisetal05b,Mininni06, Mininni08} has claimed to verify 
the importance of {\it nonlocal} triadic interactions.} Their DNS results at resolutions up to $2048^3$ 
support earlier observations of ``local transfer by nonlocal triads'' \cite{BrasseurCorrsin,
YeungBrasseur, DomaradzkiRogallo,OhkitaniKida}. This really should be called {\it ultralocal 
transfer} because it is between wavenumber modes $k,p$ differing by a fixed amount $p-k=k_0,$
where $k_0\approx 1/L$ is the low wavenumber at the energy peak. This is not to be 
confused with the traditional idea of local transfer, in which the wavenumber {\it ratio} $p/k$ 
is fixed and of order $2.$
 
How does one explain the apparent discrepancy between these various DNS results 
and the theoretical analysis in Eyink (2005)\cite{Eyink05}? At least three possibilities 
come to mind. One is that there may be an essential difference between transfers for 
band-pass filtered fields and energy flux for low-pass filtered fields. A second possibility is 
that the key difference is between sharp spectral and graded filters. Finally, a third possibility 
is that the important difference is between wavenumber bands of constant size on a linear scale 
versus those with constant size on a logarithmic scale. We stress that the rigorous results in 
Eyink (2005)\cite{Eyink05} were proved for graded low-pass filters with logarithmic binning 
(and logarithmic grading of the filter kernel). The logarithmic wavenumber scale is necessary 
in order to achieve simultaneous resolution of the dynamics both in space and in scale.     

In this paper we shall address the first possibility, by extending the rigorous results of Eyink
(2005)\cite{Eyink05}  to a band-pass decomposition via dilated, smooth filters. As in the previous
low-pass analysis, we shall show that there are simple equations that describe the  
dynamics of the band-pass energy densities, all terms of which have intuitive physical
interpretations. Furthermore, we show that band-to-band transfers are local in both scale
and in space by deriving rigorous upper bounds on the nonlocal triadic contributions. 
We shall also explain on physical grounds the difference between the rigorous upper bounds 
and the more refined scaling predictions  of Kraichnan \cite{Kraichnan66,Kraichnan71}. 
In essence, the contribution of the nonlocal triads is smaller than implied by those bounds 
due to {\it decorrelation} of disparate scales, which implies substantial cancellations of signed contributions to transfer. This idea was already proposed by Tennekes \& Lumley\cite{TL} 
(Section 8.2) and Kraichnan (1971)\cite{Kraichnan71}. We shall demonstrate these 
cancellation effects with data from a $512^3$ DNS of Navier-Stokes turbulence. 
These results taken all together confirm the conclusions of  Eyink (2005)\cite{Eyink05} that 
turbulent energy cascade is a scale-local process, in agreement with the traditional view.
Furthermore, they show that analysis of band-pass filtered energies give results essentially 
the same as the analysis of low-pass filtered energies. In the second part of this study 
\cite{AluieEyink} (hereafter referred to as II) we shall consider the remaining possibilties, 
related to sharp-spectral versus graded filters and linear versus logarithmic scale-resolution. 
These turn out to be more subtle and difficult issues to explicate.   

The outline of the paper is as follows. In section II we present our coarse-graining
approach.
Section II.A gathers some material on low-pass filtered 
fields and their energy budgets. Section II.B applies these techniques to develop
a band-pass decomposition of the kinetic energy. 
In section III we establish the scale-locality of transfer in this
decomposition. Section III.A 
establishes rigorous upper bounds on non-local transfers.  
Section III.B discusses decorrelation effects that lead to reduced 
non-local contributions to mean flux. Finally, section IV states our conclusions
and two appendices provide some mathematical details. 

\section{Band-Pass Filtering via Smooth Coarse-Graining}

\subsection{The Low-Pass Filter Technique}

The filtering approach by smooth coarse-graining of the velocity field is standard 
in the LES literature. See, for example, papers\cite{Germano92,Eyink05} and references 
therein. Any real-valued function $G(\br)$ may be chosen as a filter kernel, as long as it is 
sufficiently smooth, decays  sufficiently rapidly for large $r,$ and is normalized so that 
$ \int d^dr \,G(\br)=1. $  
It is assumed furthermore that $G$ satisfies 
$ \int d^dr \,\br \,G(\br)=\bzed$ and  
$\int d^dr \,|\br|^2G(\br)={\cal O}(1), $
so that the main support of $G$ is in the ball of radius 1 about the origin. Its dilation in $d$ 
dimensions $G_\ell (\br) = \ell^{-d} G(\br/\ell)$ shares these properties, except that it is 
supported mainly in the ball of radius $\ell.$  One then defines the ``coarse-grained'' or 
(low-pass) filtered velocity field at length-scale $\ell$ by 
\be \ol{\bu}_\ell(\bx) = \int d\br ~G_\ell(\br) \bu(\bx + \br).   
\lb{filter} \ee
In order to interpret this as a local space-average, $G$ must also be positive, or 
$G(\br)\geq 0$ for all $\br.$ It is often convenient to choose $G$ satisfying other 
desirable  properties, such as radial symmetry. Note, for example, that $G(\br)$ may be 
chosen so that both it and its Fourier transform $\widehat{G}(\bk)$ are positive and 
infinitely differentiable, with also $\widehat{G}(\bk)$ compactly supported inside the 
ball of radius 1 in Fourier space and with $G(\br)$ decaying faster than any power as 
$r\rightarrow\infty. $ See Appendix 1. 
 
By applying the filtering operation to the incompressible Navier-Stokes equation, one
obtains an evolution equation for the coarse-grained velocity $\ol{\bu}_\ell$:
\be \partial_{t} \ol{\bu}_\ell +  (\ol{\bu}_\ell\cdot\grad)\ol{\bu}_\ell = -\grad\ol{p}_\ell -
        \grad\cdot\ol{\tau}_\ell(\bu,\bu)+\nu\bigtriangleup \ol{\bu}_\ell, \,\,\,\,\,\,\,\,\,\,\,\,\,\,\,
      \grad\cdot\ol{\bu}_\ell=0  \lb{LSNS-eq} \ee
This is the same as the Navier-Stokes equation itself but with an additional
contribution from the {\it turbulent (Reynolds) stress},
\be \ol{\tau}_\ell(\bu,\bu) \equiv \ol{(\bu\bu)}_\ell - \ol{\bu}_{\ell}\ol{\bu}_{\ell}. \lb{stress}\ee  
In the rest of the paper, we shall take the liberty of dropping the $\ell$ sub-script whenever 
there is no risk of ambiguity and use the shorthand $\ol{\btau}=\ol{\tau}(\bu,\bu)$ to
denote the stress (\ref{stress}). An energy budget for the large scales may be easily 
derived from (\ref{LSNS-eq}), as:
\be \partial_{t} (\frac{1}{2}|\fu|^{2}) + \partial_{j}\bigg[(\frac{1}{2}|\fu|^{2} 
+ \ol{p})\ol{u}_{j} + \ol{u}_i\ol{\tau}_{ij} -\nu\partial_j(\frac{1}{2}|\fu|^{2})\bigg] 
  = -\ol{\Pi} -\nu |\grad\ol{\bu}|^2
 \lb{u-large} \ee
where 
\be \ol{\Pi} \equiv -(\partial_{j}\ol{u}_{i})\ol{\tau}_{ij} \lb{flux} \ee
is what is usually called the sub-grid scale (SGS) dissipation or the SGS energy flux.  
It acts as a sink term in (\ref{u-large}),  representing the energy transferred from scales larger
than $\ell$ to the smaller (sub-grid) scales at point $\bx$ in the flow.  The terms inside the 
divergence represent energy transported in space.

A small-scale energy or ``subgrid kinetic energy''  may be defined as
\be   \ol{k}\equiv \frac{1}{2}\ol{\tau}(\bu ;\bu)\equiv \frac{1}{2}\ol{\tau}(u_i,u_i). \lb{band-k} \ee
It is a positive quantity at every point in the flow if and only if the filtering kernel $G(\br)$ is positive 
for all $\br$, as was proved by Vreman et al. (1994) \cite{Vremanetal94}. Indeed, it can be rewritten 
as $\int d\br~G(\br) \frac{1}{2}|\bu(\bx+\br) - \ol{\bu}(\bx)|^2,$ which is the energy density averaged 
over a region of size $\ell$ around $\bx$ in a frame co-moving with the local large-scale velocity 
$\ol{\bu}(\bx)$  \cite{Vremanetal94}. Furthermore, integrating $\frac{1}{2}\ol{\tau}(u_i,u_i) $ in space 
gives $\int d\bx~ \frac{1}{2} |\bu(\bx)|^2- \int d\bx~ \frac{1}{2} |\ol{\bu}(\bx)|^2$, which is the total energy 
less the energy at large scales. It is easy to derive the energy budget of the small scales as:
\be \partial_t \frac{1}{2}\ol{\tau}(u_i,u_i) + \partial_{j} \bigg[ \frac{1}{2}\ol{\tau}(u_i,u_i)  
\ol{u}_j + \ol{\tau}(p,u_j) + \frac{1}{2}\ol{\tau}(u_i,u_i,u_j) -\nu\partial_j \frac{1}{2}\ol{\tau}(u_i,u_i) \bigg] 
 = \ol{\Pi}-\nu\ol{\tau}(\partial_iu_j,\partial_iu_j).
\lb{u-small} \ee
For example, see Germano (1992)\cite{Germano92}.  The energy flux $\ol{\Pi}$ now acts 
as source, representing the energy gained by the small scales from scales larger than 
$\ell$ at point $\bx$. 

\subsection{The Smooth Band-Pass Approach}

Following the same ideas, we may define a band-passed kinetic energy density 
at length-scales between $\wt{\ell}$ and $\ol{\ell}$, where 
$\wt{\ell}> \ol{\ell}$, as 
\be k_{[\ol{\ell},\wt{\ell}]}\equiv\frac{1}{2}\wt{\tau}(\ol{\bu};\ol{\bu}) =  \frac{1}{2}\wt{|\ol{\bu}|^2} 
- \frac{1}{2}|\wt{\ol{\bu}}|^2
\lb{band-energy} \ee
for filter functions $\wt{G}=G_{\wt{\ell}}$ and $\ol{G}=G_{\ol{\ell}}.$ The band-passed 
energy (\ref{band-energy}) is a straightforward generalization of the small-scale 
kinetic energy presented in the previous sub-section. Positivity again holds,  
$ \frac{1}{2}\wt{\tau}(\ol{\bu};\ol{\bu})(\bx) \ge 0,$ locally for all $\bx$ when $G(\br)\ge 0$. 
Globally, this band-pass energy is 
\begin{eqnarray*}  
\int d^dx~ \frac{1}{2} \wt{\tau}(\ol{\bu};,\ol{\bu}) & = & 
\int d^dx~ \frac{1}{2} |\ol{\bu}(\bx)|^2 - \int d^dx~ \frac{1}{2} |\wt{\ol{\bu}}(\bx)|^2 \cr
 & = & 
\frac{1}{(2\pi)^d} \int d^dk~ [1-|\widehat{G}(\bk\wt{\ell})|^2] |\widehat{G}(\bk\ol{\ell})|^2
  \frac{1}{2} |\widehat{\bu}(\bk)|^2,  
\end{eqnarray*} 
by the Plancherel identity. This is clearly a reasonable representation of the energy 
in the wavenumber range $[1/\wt{\ell},1/\ol{\ell}]$. If a sharp-spectral filter with 
$\widehat{G}(\bk)=\theta(2\pi-k)$ is used,  then this band-pass energy reduces to 
$\frac{1}{(2\pi)^d}\int_{\{2\pi/\wt{\ell}<|\bk|<2\pi/\ol{\ell}\}} d^d  k~ \frac{1}{2} |\widehat{\bu}(\bk)|^2 $,  
which is the quantity  considered in recent numerical studies 
\cite{Alexakisetal05b,Mininni06, Mininni08,Domaradzki07a,Domaradzki07b,Domaradzki09}. 

One of the great advantages of the present approach is that very simple dynamical
equations can be written for the band-pass energy densities above, which resolve 
the relevant physical process both in space and in scale. It is straightforward to show that  
\begin{eqnarray}
&& \partial_t\left({{1}\over{2}}\wt{\tau}(\ol{u}_i,\ol{u}_i)\right)  + \partial_k\left[ 
        {{1}\over{2}}\wt{\tau}(\ol{u}_i,\ol{u}_i)\wt{\ol{u}}_k 
        + \wt{\tau}(\ol{p},\ol{u}_k)  \right. \cr
&&  \,\,\,\,\,\,\,\,\,\,\,\,\,\,\,\,\,\,\,\,\,\,\,\,\,\,\,\,\,\,\,\,\,\,\,\,\,\,\,\,\,\,\,\,\,\,\,\,\,\,\,\,\,\,\,\,\,\,\,\,\,\,\,\,\,\,\,\,\,\,\,\,\,\,
       \left. +{{1}\over{2}}\wt{\tau}(\ol{u}_i,\ol{u}_i,\ol{u}_k)
+\wt{\tau}(\ol{\tau}(u_i,u_k),\ol{u}_i)-\nu\partial_k {{1}\over{2}}\wt{\tau}(\ol{u}_i,\ol{u}_i) \right] \cr
&&   \,\,\,\,\,\,\,\,\,\,\,\,\,\,\,\,\,\,\,\,\,\,\,\,\,\,\,\,\,\,\,\,\,\,\,\,\,\,\,\,\,\,\,\,\,\,\, 
= -\wt{\ol{S}}_{ij}\wt{\ol{\tau}}(u_i,u_j) + \widetilde{\ol{\tau}(u_i,u_j)\ol{S}_{ij}}
                -\nu \wt{\tau}(\partial_i\ol{u}_j,\partial_i\ol{u}_j). \lb{u-band} 
\end{eqnarray}                
If the total energy $(1/2)\int |\bu|^2$ remains finite in the limit $\nu\rightarrow 0$ 
(as expected), then the viscous terms are easily seen to be negligible in a fixed 
band $[\ol{\ell},\wt{\ell}]$ for small $\nu.$  The terms inside the space gradient
once again represent space-transport of energy. The non-viscous terms on the 
righthand side of (\ref{u-band}) represent the nonlinear transfer into and out of the band.
Note that double overlining $\wt{\ol{(\cdot)}}$ is associated to 
the convolved filter $\wt{\ol{G}}=\wt{G}*\ol{G},$ which has a length-scale 
$\wt{\ol{\ell}}\approx \wt{\ell}$ for $\ol{\ell}\ll\wt{\ell}.$ Thus, the transfer terms have 
the meaning of a flux $\wt{\ol{\Pi}}$ into scales smaller than $\wt{\ell}$ and a reverse flux
$\wt{\left(-\ol{\Pi}\right)}$ out of scales smaller than $\ol{\ell}.$

This is a good point to remark that the flux terms in eqs. (\ref{u-large}), (\ref{u-small}), and 
(\ref{u-band}) have all been defined in a Galilean-invariant way, due to the subtracted
mean terms in the definition (\ref{stress}) of the stress. Other definitions of an 
``energy flux'' are possible, such as the ``unsubtracted flux''
$$    \ol{\Pi}^{uns} \equiv -\partial_{j}\ol{u}_{i} \,\ol{u_iu_j} = \ol{\Pi}
        -\partial_j(\frac{1}{2}\ol{u}_j|\ol{\bu}|^2). $$
This differs from the standard SGS flux $\ol{\Pi}$ by a total gradient, which can be included 
in the space-transport term. This ``unsubtracted flux'' is, in fact, often employed in 
literature that considers  the sharp-spectral filter. Unfortunately, this definition is 
not pointwise Galilean-invariant, so that the amount of  ``energy cascade'' at any point 
in the fluid according to this definition would differ for observers moving at different uniform 
velocities!  Kraichnan \cite{Kraichnan64}, Speziale \cite{Speziale} and Germano \cite{Germano92} 
have all emphasized the importance of Galilean invariance.  Our definitions are the unique 
ones which preserve the pointwise Galilean-invariance of energy flux. There are 
non-Galilean-invariant terms in our balance equations,  but,  as is physically natural, 
they are all associated to space-transport of kinetic energy. 

The previous methods may be used to introduce a band-pass decomposition of 
the energy at a geometric sequence of scales $\ell_n=\rho^{-n}L$ with $\rho>1.$ 
The result is easiest to state for the special case of filter kernels $G$ which satisfy
the ``S-type'', or sharp-spectral-like, condition that 
\be  \widehat{G}(\bk) = \left\{ \begin{array}{ll}
                                            1 & \mbox{if $|\bk|<1$} \cr
                                             0 & \mbox{if $|\bk|>\rho$} \cr
                                             \end{array}\right.  .  \lb{S-type} \ee
As we show in Appendix 1, it is possible to construct kernels of this type for which
$\widehat{G}(\bk)$ is $C^\infty$ and thus $G(\br)$ decays faster than any power 
of $r$ as $r\rightarrow\infty.$ We introduce $G_n(\br) = \ell_n^{-d} G(\br/\ell_n)$
and the corresponding low-pass filter $\ol{f}_n= G_n*f$ for any space function $f(\bx).$
It is then possible to show that 
\be {{1}\over{2}} \int |\bu|^2= {{1}\over{2}}\int |\ol{\bu}_{0}|^2
   + {{1}\over{2}}\sum_{n=1}^N \int \tau_{n-1}(\ol{\bu}_n;\ol{\bu}_n) + 
      {{1}\over{2}}\int \tau_N(\bu;\bu). \lb{band-decomp} \ee
for any integer $N\geq 1.$ For the proof of this formula, and also a more general result 
without condition (\ref{S-type}), see Appendix 2.  It is easy to see that series
(\ref{band-decomp}) converges as $N\rightarrow\infty$ whenever $(1/2)\int |\bu|^2<\infty.$     
      
Using (\ref{u-large}),(\ref{u-band}),(\ref{u-small}) one can infer evolution equations 
for the total band energies:
$$ (d/dt)\int {{1}\over{2}}|\ol{\bu}_{0}|^2 = -\int \Pi_0, $$
$$ (d/dt)\int {{1}\over{2}}\tau_{n-1}(\ol{\bu}_n;\ol{\bu}_n) = \int \Pi_{n-1}-\int \Pi_n \,\,\,\,\,\,\,\,
       {\rm for} \,\,\,\,n=1,...,N $$
\textcolor{black}       
{$$ (d/dt)\int {{1}\over{2}}\tau_N(\ol{\bu};\ol{\bu})=\int \Pi_N -\nu\int \tau_N(u_{i,j},u_{i,j}). $$
Note that viscous terms may be neglected as $\nu\rightarrow 0$ in all of these equations
except the final one, for $n=N,$ where it balances the flux from the larger scales. If the 
turbulence is driven by a body force at length-scales $>L,$ then an additional source term 
would appear in the first equation for $n=0.$ This hierarchy makes apparent the stepwise
nature of the cascade, if each of the flux terms $\Pi_n=-\grad\ol{\bu}_n\,\bdots\,
\tau_n(\bu,\bu)$ for inertial-range length-scales $\ell_n$ depends only upon velocity 
modes of comparable scale. }

\textcolor{black}
{In a recent paper, Cheskidov et al. (2008)\cite{Cheskidovetal08} have studied 
turbulent cascade dynamics by a band-pass decomposition of the velocity field 
with graded filters similar to ours. They derived thereby a number of estimates 
related to scale-locality of energy transfer. However, as noted by 
Domaradzki \& Carati (2007)\cite{Domaradzki07a} the naive decomposition 
of the kinetic energy following from such a decomposition of the velocity field 
in general will contain off-diagonal terms:
$$  {{1}\over{2}} \int |\bu|^2= {{1}\over{2}}\int |\ol{\bu}_{0}|^2
   + {{1}\over{2}}\sum_{n,m=1}^\infty \int \ol{\bu}_n\bdot \ol{\bu}_m. $$
In the approach of Cheskidov et al. (2008)\cite{Cheskidovetal08} the summation 
is limited to indices with $|n-m|\leq 1$ but still retains such off-diagonal terms of
indefinite sign.  It is thus not clear in this approach how precisely to identify the kinetic 
energy at a given length-scale. By contrast, our decomposition (\ref{band-decomp}) 
of kinetic energy contains only non-negative, diagonal terms.  The complications with the 
naive band-pass decomposition lead to related difficulties in the analysis of scale transfers 
of energy. For example,  the ``band-flux'' defined in eq.(27) of Cheskidov et al. (2008)
\cite{Cheskidovetal08} is not the difference of the usual energy fluxes into and out of the band, 
as we found above in our approach, but contains an additional term [see their eq.(30)]
with unclear physical interpretation,} \textcolor{black}{which those authors had to estimate.}

 
\section{Locality of Energy Cascade}

One of the advantages of our approach to band-pass decomposition of kinetic 
energy is that the proofs of scale-locality of Eyink (2005)\cite{Eyink05} directly apply. 
In fact, the flux quantities $\Pi_n$ were exactly the objects shown in that work to be local! 
\textcolor{black}{
To make connection with standard spectral approaches, we define the {\it total transfer}
into the $n$th band by 
$$ {\rm T}_n(\bu,\bu,\bu)\equiv \Pi_{n-1}(\bu,\bu,\bu)-\Pi_n(\bu,\bu,\bu). $$ 
Clearly, this quantity must be dominated by the contributions of local triads if 
this is true separately for both $\Pi_{n-1}$ and $\Pi_n$. The total spectral transfer is often 
further decomposed as ${\rm T}_n=\sum_p {\rm T}_{n,p}= \sum_{m,p} {\rm T}_{n,m,p}$
into triadic contributions from distinct wavenumber bands. If we do so in the present 
approach  of smooth coarse-graining, then it follows that transfer for non-local modal triads  
is negligibly small.}

It should be emphasized that there  {\it are} non-local triadic interactions in turbulent 
dynamics. For example, non-local interactions dominate in the space-transport 
of energy by convective sweeping. However, space-transport effects  disappear 
upon integration over space and make no net contribution to the transfer of energy 
between scales.  On the other hand, the energy fluxes $\Pi_n$ which are responsible 
for the inter-scale exchange of energy are scale-local,  under well-defined conditions that exist 
in the inertial-range of turbulent flows.  We give below a brief summary of the demonstration
from Eyink (2005)\cite{Eyink05}. 

\subsection{The Proof of Scale-Locality}

The energy flux $\ol{\Pi}_\ell$ is a triplet quantity that depends upon three velocity modes. 
This dependence may be made explicit by writing 
 $$ \ol{\Pi}_\ell(\bu,\bv,\bw) \equiv -\grad\ol{\bu}_\ell: \tau_\ell(\bv,\bw). $$ 
(In the discussion of locality it is important to indicate the scale $\ell,$ which we do in this whole
section). We say that the energy flux $\ol{\Pi}_\ell=\Pi_\ell(\bu,\bu,\bu)$ is \emph{infrared local} 
if the contribution from $\bu$ at scales much larger than $\ell$ is negligible. In other words,
if $\bu$ in the three arguments $(\bu,\bv,\bw)$, \emph{or any subset thereof}, were to be replaced 
by $\ol{\bu}_\Delta,$ then the result would be $\ll \ol{\Pi}_\ell$ for $\Delta\gg \ell.$ We can define 
a corresponding concept by employing the ``small-scale'' or high-pass filtered field 
\be \bu_\delta' \equiv \bu-\ol{\bu}_\delta,  \lb{lLP-u} \ee
which contains no modes at scales $>\delta.$ We say that $\ol{\Pi}_\ell=\ol{\Pi}_\ell(\bu,\bu,\bu)$ 
is \emph{ultraviolet local} if replacing $\bu$ by $\bu'_\delta$ in  the three arguments $(\bu,\bv,\bw)$, 
\emph{or any subset thereof},  were to give a result $\ll \ol{\Pi}_\ell$ whenever $\delta\ll \ell.$
This means that the contribution to energy flux from modes at scales much smaller than $\ell$ 
is negligible. If a quantity is both infrared and ultraviolet local, then we say that it is (scale-)local. 

They key to proving scale-locality of the energy flux is the observation that all of 
the quantities $\grad\ol{\bu}_\ell,\tau_\ell(\bu,\bu),$ and $\bu_\ell'$ can be written 
entirely in terms of  \emph{velocity-increments}:
\be \delta\bu(\bx; \br) \equiv \bu(\bx + \br) - \bu(\bx) 
\lb{increment} \ee
for separation distances $|\br|<\ell$ (or some moderate multiple of $\ell$) and do not 
depend  upon the absolute velocity $\bu(\bx)$ itself. Heuristically, 
\be \grad\ol{\bu}_\ell \sim {\mathcal O}(\delta u(\ell)/\ell) 
\lb{strain-increment}\ee
\be \ol{\btau}_\ell(\bu,\bu) \sim {\mathcal O}(\delta u^2(\ell)) 
\lb{stress-increment}\ee
\be \bu_\ell' \sim {\mathcal O}(\delta u(\ell)) 
\lb{smallu-increment}\ee
The symbol $\sim{\mathcal O}$ stands for  ``same order-of-magnitude as.'' 
For rigorous details, see Eyink (2005)\cite{Eyink05}. 

It thus becomes \textcolor{black}{sufficient to} show that  velocity-increments themselves are scale-local.
This is not generally true, for arbitrary solutions of the Navier-Stokes equations, but it is true 
under conditions that exist in the inertial-range of turbulent flows. The velocity fields  in 
such turbulent solutions are not space-differentiable but only \emph{H\"{o}lder continuous}
\cite{Frisch}. A field $\bu$ is said to be H\"{o}lder continuous at $\bx$ with exponent $0<h<1$ 
if its increments satisfy a rigorous big-$O$ upper bound 
\be |\delta\bu(\bx;\br)| =O(r^h). \lb{hoelder} \ee
The (maximal) H\"{o}lder exponent at point $\bx$ is the largest value $h$ such that
the inequality (\ref{hoelder}) still holds.  We now observe that infrared contribution to the increment, 
$\delta \ol{\bu}_\Delta (\bx;\br)  = \ol{\bu}_\Delta (\bx+ \br) - \ol{\bu}_\Delta (\bx)$, is negligible 
if $h<1$. Since the low-pass filtered field $\ol{\bu}_\Delta$ is very smooth, 
we can approximate it well by the leading term in its Taylor expansion: 
\be  \delta \ol{\bu}_\Delta (\br;\bx) \approx \br\bdot\grad\ol{\bu}_\Delta(\bx) 
      \sim {\mathcal O}\left(r  \frac{\delta u(\Delta)}{\Delta}\right)=O(r\Delta ^{h-1}). \lb{UV-locality} \ee
Compared with the full increment $\delta \bu(\br)\sim {\mathcal O}(r^h),$ the large-scale contribution
is smaller by a factor of $O((r/\Delta)^{1-h}),$ which is indeed negligible if $\Delta \gg r$ and $h<1.$
We next observe that the ultraviolet contribution to the increment $\delta \bu'_\delta (\bx;\br) = 
\bu'_\delta (\bx+\br) - \bu^{'}_\delta (\bx)$ is negligible if $h>0.$ This is even easier to see, 
since, from (\ref{smallu-increment}), the high-pass filtered field itself is small:  
$\bu'_\delta \sim {\mathcal O}(\delta u(\delta))$. In that case,
\be \delta\bu'_\delta (\bx;\br) \sim {\mathcal O}(\delta u(\delta))=O(\delta^h). \lb{IR-locality} \ee
Compared with the full increment, the small-scale contribution is smaller by a factor of 
$O((\delta/r)^h),$ which is also negligible if $\delta\ll r$ and $h>0.$ All of these results hold 
pointwise, locally in space \cite{Eyink05}. 

One difficulty with the above pointwise analysis is that inequality (\ref{hoelder}) does not hold as 
a scaling law \cite{FrischVergassola91,Arneodoetal95}.  The upper bound in (\ref{UV-locality}) 
is vanishing for $\Delta\rightarrow\infty$ and the bound in (\ref{IR-locality}) is vanishing for 
$\delta\rightarrow 0.$ However, because H\"older continuity does not correspond to pointwise 
scaling, one cannot make precise statements about the {\it fractional contributions} from scales
$>\Delta$ or $<\delta.$ On the other hand,  it is well-known that there {\it is} scaling of $p$th 
structure functions of velocity increments in the turbulent inertial-range. That is, 
$\langle|\delta\bu(\br)|^p\rangle\sim {\mathcal O}(u^p_{rms} (r/L)^{\zeta_p}) $
where $\langle\cdot\rangle=(1/V)\int_V (\cdot)$ denotes volume-average. A global analogue
of (\ref{hoelder}) thus holds for the $L_p$-norm $\|\cdot\|_p=\big\langle\,|\cdot|^p\big\rangle^{1/p}$ as a 
true scaling law, i.e.  for some dimensionless constant $A_p$
\be \|\delta\bu(\br)\|_p\sim u_{rms} A_p (r/L)^{\sigma_p} \lb{besov} \ee
with $\sigma_p=\zeta_p/p.$ Experiments and simulations indicate that $0<\sigma_p<1,$ 
and, in particular, $\sigma_3\doteq 1/3.$ The relation (\ref{besov}) corresponds to 
{\it Besov regularity} \cite{Eyink95,BasdevantPerrier96} in place of the H\"older continuity
relation (\ref{hoelder}). It is not difficult to derive global $L_p$-mean analogues of 
IR-locality of velocity-increments
\be  \|\delta \ol{\bu}_\Delta (\br)\|_p =O(r\Delta ^{\sigma_p-1}). \lb{UV-locality-Lp} \ee
and UV-locality of velocity-increments
\be \|\delta\bu'_\delta (\br)\|_p =O(\delta^{\sigma_p}). \lb{IR-locality-Lp} \ee
For details, see Eyink (2005)\cite{Eyink05}. Together with (\ref{besov}) we may make the precise 
statement that the modes $>\Delta$ give a fractional contribution $O((\ell/\Delta)^{1-\sigma_p})$
and the modes $<\delta$ a fractional contribution $O((\delta/\ell)^{\sigma_p}).$

These results for velocity-increments imply the locality of the energy flux. Here we shall 
just state two concrete estimates which correspond closely to classical statements of 
locality in the literature. First we consider the contribution to the flux from a large-scale
strain, or $\Pi_\ell(\ol{\bu}_\Delta,\bu,\bu)=-\grad\ol{\ol{\bu}}_{\Delta,\ell}:\tau_\ell(\bu,\bu).$
Using the H\"older inequality $|\langle fgh\rangle|\leq \|f\|_3\|g\|_3\|h\|_3$ it is easy to show
from the foregoing that 
\be \langle\Pi_\ell(\ol{\bu}_\Delta,\bu,\bu)\rangle=O(\varepsilon (\ell/\Delta)^{1-\sigma_3}). 
\lb{IR-flux} \ee
We have used the relation $\varepsilon\sim u_{rms}^3/L$ in order to introduce the mean flux 
$\varepsilon.$ Next we consider the contribution from the small-scale stress, or 
$\Pi_\ell(\bu,\bu_\delta',\bu_\delta')=-\grad\ol{\bu}_\ell:\tau_\ell(\bu_\delta',\bu_\delta').$
Note that {\it both} of the modes in the stress must be small-scale together, if the filter 
kernel $G$ is ``S-type'', because of wavenumber conservation constraints. Using again 
the H\"older inequality and the previous estimates on increments, one finds that
\be \langle\Pi_\ell(\bu,\bu_\delta',\bu_\delta')\rangle=O(\varepsilon (\delta/\ell)^{2\sigma_3}) 
\lb{UV-flux} \ee
\textcolor{black}{
In deriving (\ref{IR-flux}),(\ref{UV-flux}) it was assumed that $\sigma_3\doteq 1/3,$ so that 
the exponents in both of the bounds are very close to $2/3.$ Also, to conclude that the 
non-local flux contributions in (\ref{IR-flux}) and (\ref{UV-flux}) are much smaller than 
$\langle\Pi_\ell\rangle$ for $\Delta\gg \ell$ and $\delta\ll\ell,$ it must be assumed that 
the mean flux is non-zero. Of course, this is a completely realistic assumption in a
constant-flux inertial range.}

\subsection{Decorrelation Effects}

At first sight, the results stated in (\ref{IR-flux}) and (\ref{UV-flux}) contradict the predictions 
of Kraichnan \cite{Kraichnan66,Kraichnan71} using closure calculations, which would yield
a larger exponent $4/3$ instead of $2/3.$ However, one must keep in mind that our estimates
above are only {\it upper bounds} or so-called big-$O$ bounds, which crudely replace all the
velocity-increments with their absolute magnitudes. On the other hand, Kraichnan's estimates
are more refined (but non-rigorous) scaling predictions which take into account cancellations 
in the signed contributions. See the heuristic discussion in Section 1 of Kraichnan 
(1971)\cite{Kraichnan71} and the similar discussion by Tennekes and Lumley\cite{TL},  
sections 3.2 and 8.2. Their considerations can be carried over to our framework and then 
yield similar results.  We now discuss this briefly.

The easiest case to consider is infrared locality of the flux. In the average over space
$$ \langle \Pi_\ell(\ol{\bu}_\Delta,\bu,\bu)\rangle
     =-\frac{1}{V}\int_V d^dx\,\grad\ol{\ol{\bu}}_{\Delta,\ell}:\tau_\ell(\bu,\bu), $$
the velocity-gradient $\grad\ol{\ol{\bu}}_{\Delta,\ell}$ varies on a long length-scale $\sim\Delta$
whereas the stress $\tau_\ell(\bu,\bu)$ varies on the short scale $\sim \ell.$ The average is only 
nonzero because of correlations between these two quantities, when the turbulence is either
homogeneous or isotropic so that $\langle\grad\bu\rangle=0.$ An estimate of the correlation
factor proposed in \cite{TL,Kraichnan71} is that
$$ \rho(\grad\ol{\bu}_\Delta,\btau_\ell) = O(S(\Delta)/S(\ell)) $$
where $S(\ell)\sim {\mathcal O}(\delta u(\ell)/\ell)$ is the strain-rate at scale $\ell.$ This can be  
plausibly justified by appealing to ergodicity to rewrite the space-averages as time-averages
and then arguing that the correlation should be proportional to the ratio of the turnover rates
at the two length-scales. Taking decorrelation into account yields an 
improved estimate\footnote{We remark here that the additional decorrelation 
resulting from space averaging of signed quantities can explain why Benzi et al. 
\cite{Benzietal99} see a faster than expected decay rate of the multi-scale structure 
functions shown in their Figure 8.} 
$$ \langle \Pi_\ell(\ol{\bu}_\Delta,\bu,\bu)\rangle \sim
\langle\frac{\delta u(\Delta)}{\Delta}\delta u^2(\ell)\rho(\grad\ol{\bu}_\Delta,\btau_\ell)\rangle
     \sim \frac{\ell}{\Delta^2}\langle \delta u^2(\Delta)\delta u(\ell)\rangle. $$
The latter expression is a ``multiscale structure-function'' of a type much discussed in 
the literature. It can be estimated by means of a multiplicative cascade ansatz 
\cite{Kolmogorov62,Frisch} or, equivalently, by so-called ``fusion rules'' 
\cite{Eyink93,LvovProcaccia,Benzietal99}. The result is that 
$$ \langle \delta u^2(\Delta)\delta u(\ell)\rangle\sim 
      \langle \delta u^3(\Delta)\frac{\delta u(\ell)}{\delta u(\Delta)}\rangle
    \sim u_{rms}^3 \left(\frac{\Delta}{L}\right)^{\zeta_3} \left(\frac{\ell}{\Delta}\right)^{\zeta_1}.
$$
This gives an estimate for the flux contribution from the large-scale strain 
$$ \langle \Pi_\ell(\ol{\bu}_\Delta,\bu,\bu)\rangle \sim
      {\mathcal O}\left(\varepsilon \left(\frac{\ell}{\Delta}\right)^{1+\zeta_1}\right) $$
if we take $\zeta_3=1.$  The above estimate incorporates possible intermittency 
effects, but, assuming K41 scaling so that $\zeta_1=1/3,$ we recover Kraichnan's
predicted 4/3 exponent \cite{Kraichnan66,Kraichnan71}. 

A similar discussion may be given for ultraviolet locality of the flux. Note first that  
$\tau_\ell(\bu_\delta',\bu_\delta')=\overline{(\bu_\delta'\bu_\delta')_\ell}$ for small 
enough $\delta,$ if the filter kernel $G$ is ``S-type'' so that $\ol{(\bu'_\delta)_\ell}
=0$ for $\delta<\ell/\rho.$ Thus, 
$$ \langle \Pi_\ell(\bu,\bu'_\delta,\bu'_\delta)\rangle 
     =-\frac{1}{V}\int_V d^dx\,\grad\ol{\ol{\bu}}_{\ell,\ell}: \bu_\delta'\bu_\delta' $$
Since the velocity-gradient $\grad\ol{\ol{\bu}}_{\ell,\ell}$ varies on a long length-scale 
$\sim\ell$ whereas the factors $\bu_\delta'$ vary on the short scale $\sim \delta,$ 
we can expect decorrelation by a factor $S(\ell)/S(\delta).$  This gives
$$ \langle \Pi_\ell(\bu,\bu_\delta',\bu_\delta')\rangle \sim
\langle\frac{\delta u(\ell)}{\ell}\delta u^2(\delta)\rho(\grad\ol{\bu}_\ell,\btau_\delta)\rangle
     \sim \frac{\delta}{\ell^2}\langle \delta u^2(\ell)\delta u(\delta)\rangle. $$
Using the same heuristic estimation as before by the cascade ansatz or fusion-rule 
formula, one obtains the similar result  
 $$  \langle \Pi_\ell(\bu,\bu_\delta',\bu_\delta')\rangle \sim
    {\mathcal O}\left(\varepsilon \left(\frac{\delta}{\ell}\right)^{1+\zeta_1}\right).  $$
This again yields the $4/3$ scaling predicted by Kraichnan 
\cite{Kraichnan66,Kraichnan71} if we assume the K41 value $\zeta_1=1/3. $  
         
In fact, we can expect strong {\it pointwise} decorrelation effects in the ultraviolet contributions
to the energy flux, due to local averaging over the scale $\ell$ even without the global 
average over space. If the statistics of $\bu_\delta'$ were isotropic in the ball of radius 
$\sim\ell$ around point $\bx,$  then $\overline{(\bu_\delta'\bu_\delta')_\ell} = 
\frac{1}{d}\overline{(|\bu_\delta'|^2)_\ell}{\bf I}.$ In reality, the statistics will not be exactly 
isotropic, because there is a local mean velocity-gradient $\grad\ol{\bu}_\ell(\bx)$ 
which distorts the smaller scales. Thus, we expect that
$$ \overline{(\bu_\delta'\bu_\delta')_\ell} = \frac{1}{d}\overline{(|\bu_\delta'|^2)_\ell}{\bf I} 
      + {\mathcal O}(S(\ell)/S(\delta)|u'_\delta|^2), $$
based on the same plausible reasoning as Tennekes \& Lumley\cite{TL} and 
Kraichnan (1971)\cite{Kraichnan71}. The leading term does not make a contribution 
to energy flux, because of the zero trace of the velocity-gradient.  Thus, pointwise, 
we expect that 
$$  \Pi_\ell(\bu,\bu'_\delta,\bu'_\delta) = {\cal O}\left(\frac{\delta}{\ell^2}\delta u^2(\ell)\delta u(\delta)\right)
       = {\mathcal O}\left(\left(\frac{\delta}{\ell}\right)^{1+h}\right)\Pi_\ell(\bu,\bu,\bu),$$
where  $\Pi_\ell(\bu,\bu,\bu)={\mathcal O}(\delta u^3(\ell)/\ell)$ and we estimate 
$\delta u(\delta)={\mathcal O}(\delta^h)$ with local H\"older exponent $h$ at $\bx.$    
This is considerably smaller than the rigorous pointwise upper bound established 
by Eyink (2005)\cite{Eyink05}. 

\subsection{Numerical Results}

The arguments in the preceding section are plausible but non-rigorous. It is therefore 
worthwhile to investigate the predicted decorrelation effects in numerical simulations.
We present here the results of a $512^3$ DNS of Navier-Stokes turbulence, whose
details are discussed in paper II. We use Gaussian filters $G_\ell(r)=(\pi/\ell^2)^{3/2}\exp(-\pi^2 r^2/\ell^2)$ 
to scale-decompose the simulated velocity field. To facilitate comparison 
with results for spectral filters we employ an ``equivalent wavenumber'' $K$ with 
$\ell_K=2\pi/K.$

We first check the decorrelation argument in the case of IR locality, by measuring both 
the mean absolute flux $\langle|\Pi_{\ell_K}(\ol{\bu}_{\ell_Q},\bu,\bu)|\rangle$ and
the mean signed flux $\langle\Pi_{\ell_K}(\ol{\bu}_{\ell_Q},\bu,\bu)\rangle$, for $\ell_Q \ge 2\ell_K$. 
The former 
is immune to cancellation effects, whereas the second should experience decorrelation 
from space averaging. The plots in figure \ref{figIR} are consistent with a $Q^{2/3}$ 
scaling for absolute flux and a $Q^{4/3}$ scaling for signed flux, exhibiting the faster decay 
due to decorrelation effects. The fitted IR exponents are slightly smaller than those predicted, 
consistent with the results of Domaradzki \& Carati\cite{Domaradzki07a,Domaradzki09}. 

We also check the decorrelation argument for UV locality by measuring the 
absolute flux $\langle|\grad\ol{\ol{\bu}}_{\ell_K,\ell_K}: \bu_{\ell_P}'\bu_{\ell_P}' |\rangle$ 
and the signed flux $\langle\grad\ol{\ol{\bu}}_{\ell_K,\ell_K}: \bu_{\ell_P}'\bu_{\ell_P}' \rangle$,
for $\ell_P \le \ell_K /2$.
The plots in figure \ref{figUV} are again consistent with respective decays of $P^{-2/3}$ and
$P^{-4/3},$ as expected from our arguments. The fitted UV exponents are somewhat larger
than predicted, also as observed by Domaradzki \& Carati\cite{Domaradzki07a,Domaradzki09}. 

\begin{figure}
\begin{center}
\epsfig{figure= 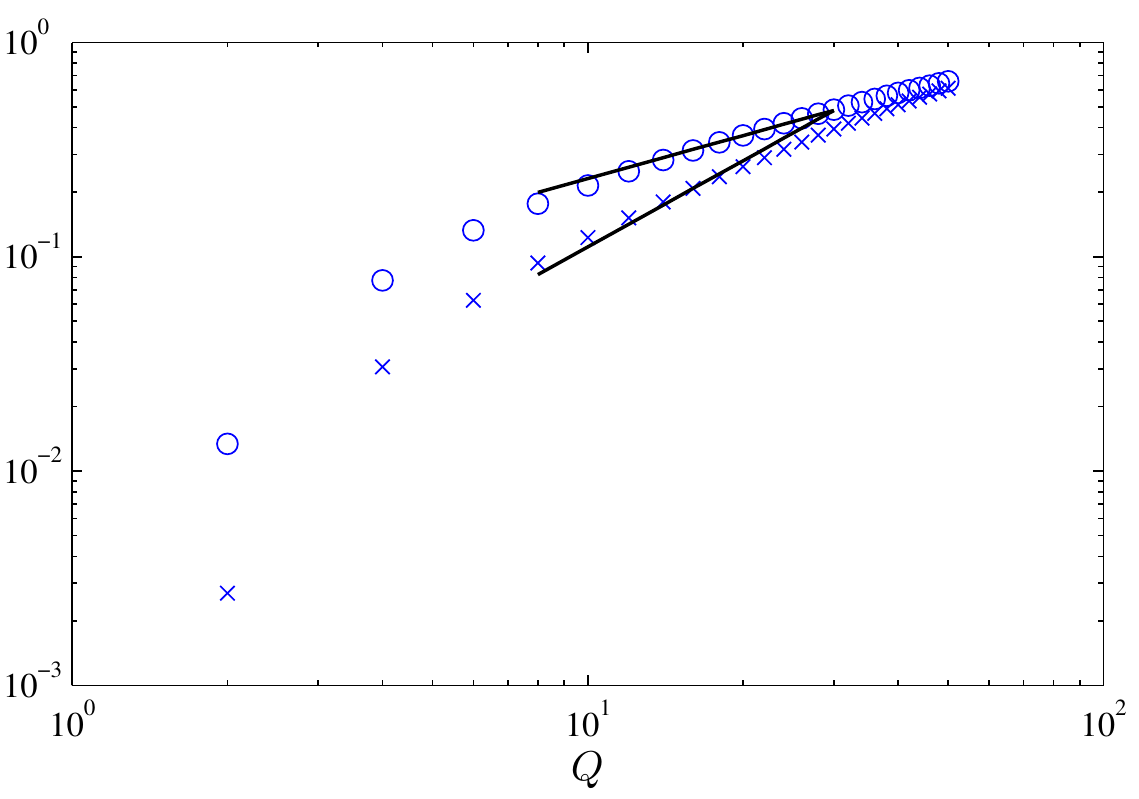}
\end{center}
 \caption{
For $K=100$, we plot as functions of $Q\le K/2$ both 
$\langle|\Pi_{\ell_K}(\ol{\bu}_{\ell_Q},\bu,\bu)|\rangle/\langle|\Pi_{\ell_K}|\rangle$ ($\circ$) 
and $\langle\Pi_{\ell_K}(\ol{\bu}_{\ell_Q},\bu,\bu)\rangle/\langle\Pi_{\ell_K}\rangle$ 
($\times$)  using a gaussian filter. Straight lines of slopes $2/3$ and $4/3$ are included
for reference. They extend over the fitting region which gives a slope $0.7$ for the absolute flux 
and $1.1$ for the signed flux. The difference in slope demonstrates the decorrelation effect 
due to space averaging.
 }
 \lb{figIR}
\end{figure}

\begin{figure}
\begin{center}
\epsfig{figure= 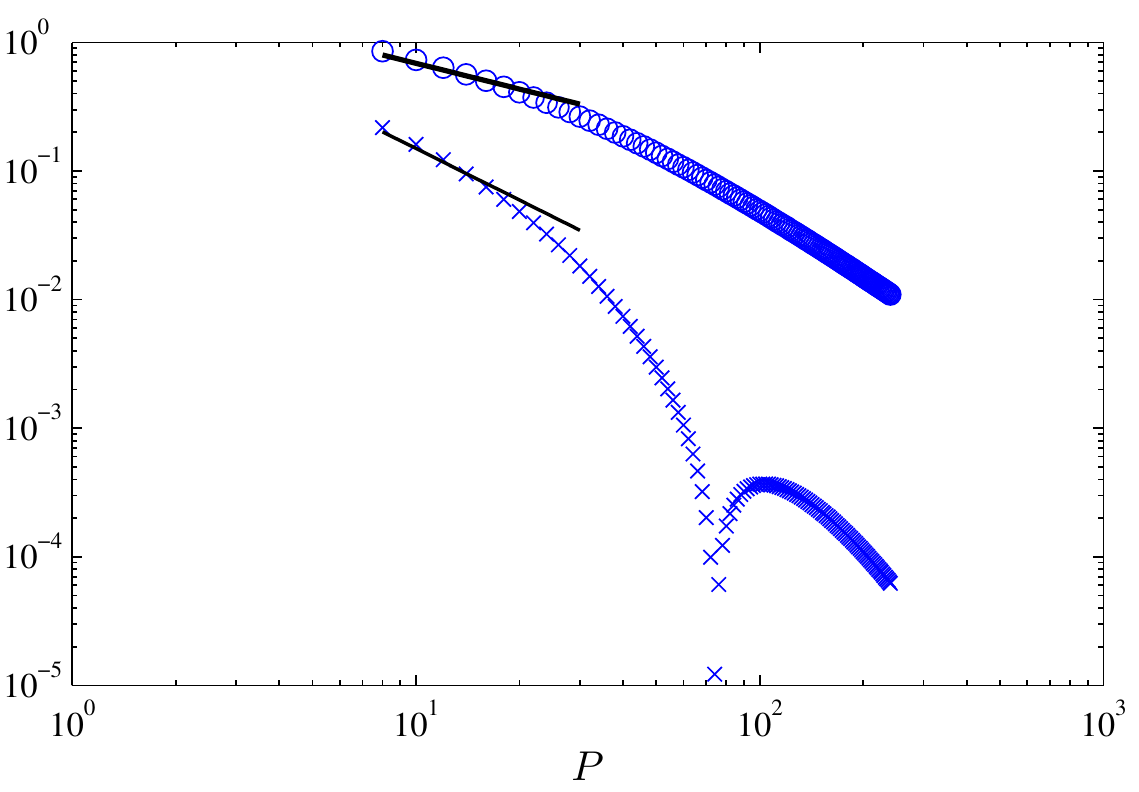}
\end{center}
 \caption{
 For $K=4$, we plot as functions of $P\ge 2K$ both $\langle|\grad\ol{\ol{\bu}}_{\ell_K,\ell_K}: 
 \bu_{\ell_P}'\bu_{\ell_P}' |\rangle/\langle|\Pi_{\ell_K}|\rangle$ ($\circ$)
 and $\langle\grad\ol{\ol{\bu}}_{\ell_K,\ell_K}: \bu_{\ell_P}'\bu_{\ell_P}' \rangle/
 \langle\Pi_{\ell_K}\rangle$  ($\times$) using a gaussian filter.  Straight lines 
 with slopes $2/3$ and $4/3$ are added for reference. They extend over the fitting region which 
 gives a slope $0.9$ for the absolute flux and $1.9$ for the signed flux, demonstrating the 
 decorrelation effect due to space averaging. The kink at very large $P$ is due to negative values 
 of the function when plotted on a log-log graph.
 }
 \lb{figUV}
\end{figure}

\newpage 

\section{Conclusions}

Let us summarize our main results. First, we have shown how to obtain a band-pass decomposition
of kinetic energy with a smooth, graded filter, corresponding to length-scales in a geometric
sequence. This decomposition permits simultaneously resolution of the turbulence dynamics 
both in scale and in physical space. Second, we have demonstrated that the interband transfers 
in this decomposition are dominated by local triads of modes, assuming only the scaling properties 
of velocity-increments that are expected (and observed) to hold in turbulent flows. Finally, we have 
explained the apparent discrepancy between our rigorous upper bounds and the scaling predictions 
of Kraichnan \cite{Kraichnan66,Kraichnan71}, as due to cancellations that appear in averages over
space. These decorrelation effects have been confirmed by results from a DNS of 
homogeneous, isotropic Navier-Stokes turbulence. 
 
We should emphasize the importance of the pointwise results on scale locality that were
deduced here and in the previous work \cite{Eyink05}. If local triadic interactions dominated 
the energy cascade only ``on average'' and not pointwise in space, for individual flow realizations,
then there would be scant basis for the usual hypothesis of small-scale universality.  Higher-order
statistics in particular could show non-universal effects from such non-local interactions, 
if they were sizable. Fortunately, this is not the case. It should be noted, however, that 
simultaneous resolution of the turbulence fields both in space and in scale requires that
our decomposition employ bands of constant width on a logarithmic scale. Our upper 
bounds on non-local contributions are exponentially small in the logarithmic band-number. 
However, this corresponds to only a slow, power-law decay in length-scale ratios,
in agreement with earlier conclusions of Kraichnan \cite{Kraichnan59,Kraichnan66} and 
Tennekes \& Lumley \cite{TL} about the ``diffuseness'' or  ``leakiness'' of the energy cascade. 
Thus, very high Reynolds numbers and long inertial ranges could be required before universal
statistics are obtained at small scales.  

A final conclusion of this paper is that smooth band-decompositions which provide 
space-scale resolution have energy transfer between bands due mainly to local triadic 
interactions. It remains to account for the disagreement with DNS studies \cite{BrasseurCorrsin,YeungBrasseur,DomaradzkiRogallo,OhkitaniKida,Alexakisetal05b,
Mininni06, Mininni08} that claim to observe transfer predominantly by non-local triads.  
This is the subject of the following Part II of this work. 
 
\vspace{.1in} 

\noindent {\small
{\bf Acknowledgements.} H.A. wishes to thank Shiyi Chen, Minping Wan and Dmitry Shapovalov 
for assistance in developing the simulation code used in this work.  Computer time was provided 
by the Digital Laboratory for Multi-Scale Science at the Johns Hopkins University. This work was 
supported by NSF grant \# ASE-0428325 at the Johns Hopkins University.}

\newpage
\setcounter{section}{0}
\renewcommand{\thesection}{Appendix \arabic{section}:}
\renewcommand{\theequation}{A-\arabic{equation}}
\setcounter{equation}{0}  
\setcounter{Prop}{0}  

\section{Filter Kernels with Compact Spectral Support}

In this appendix we explain briefly some elementary methods to construct 
filter kernels with compact spectral support, used to simplify several 
arguments in the text. 

\begin{figure}
\begin{center}
\epsfig{figure= 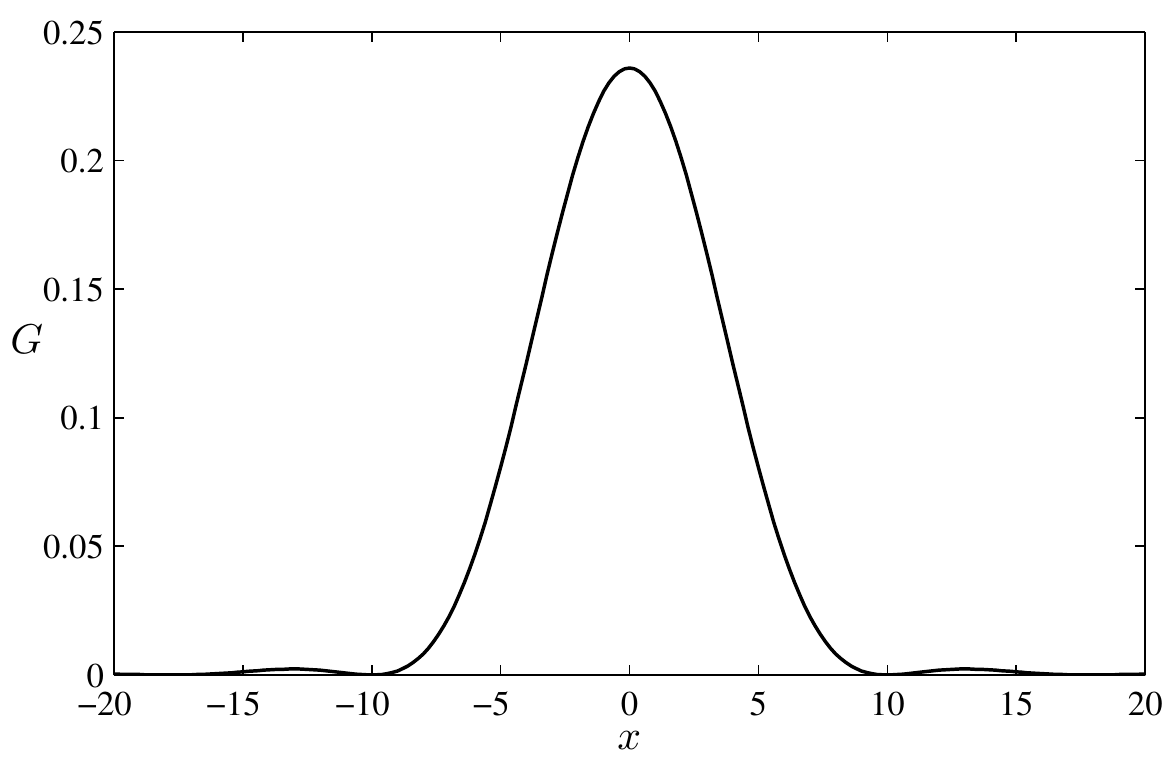}
\end{center}
\caption{A filter kernel $G(x)$ which is positive, real analytic, and decays faster than any power
at large distances. 
}\label{fig:G}
\end{figure}

\begin{figure}
\begin{center}
\epsfig{figure= 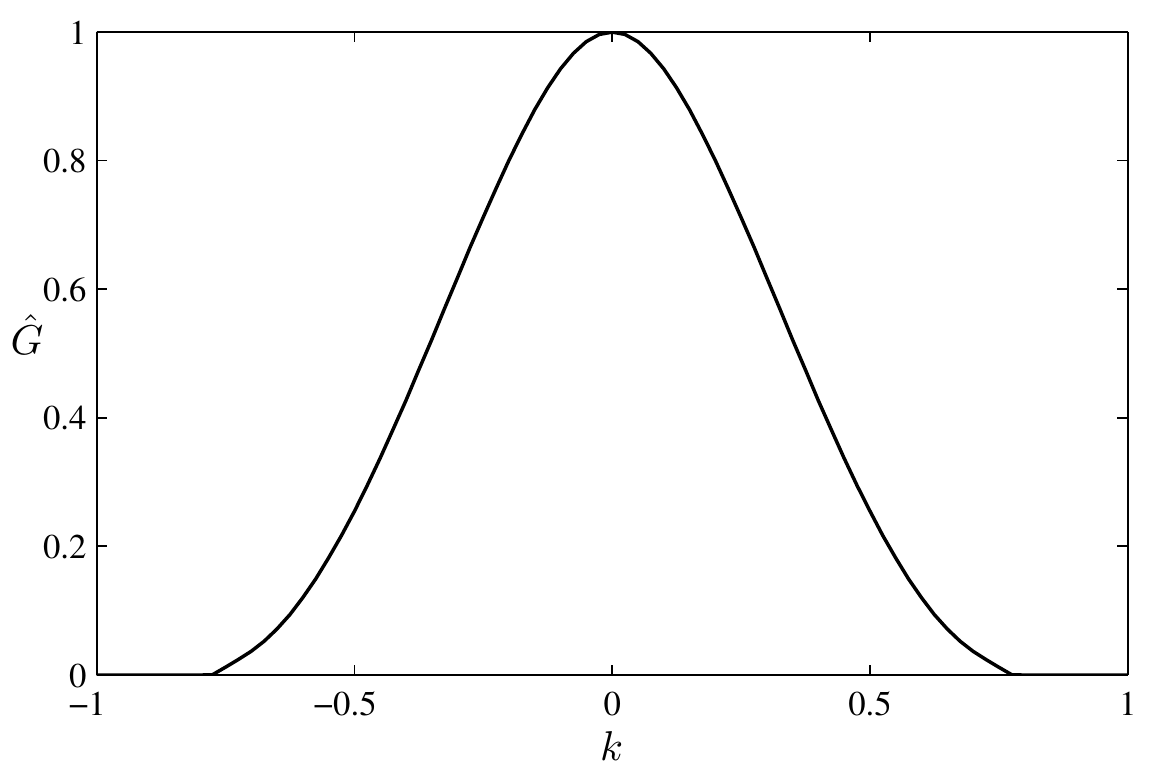}
\end{center}
 \caption{The Fourier transform $\hat{G}(k)$ of the kernel in Fig.\ref{fig:G} is positive, $C^\infty$,
and compactly supported in the unit ball.
 }\label{fig:hatG}
\end{figure}

The first construction yields a filter kernel whose Fourier transform $\widehat{G}$ is 
$C^\infty$ and supported inside the ball in Fourier space with $|\bk|<1.$ Thus, 
$G$ is decaying faster than any power of $|\br|$ as $|\br|\rightarrow\infty$ and 
is analytic in $\br$. We shall show that  both $G$ and $\widehat{G}$ may be chosen,
furthermore, to be nonnegative and rotationally symmetric. To begin the construction, 
define a function $\widehat{G}_0(\bk)$ of $k=|\bk|$ which is $C^\infty$ and supported 
inside the ball $|\bk|<1/2.$ This is a standard problem in analysis. For example, 
one may take  
$$ \widehat{G}_0(\bk) = \left\{ \begin{array}{ll}
                                                   \exp\left(-{{k^2}\over{1/4-k^2}}\right) & |\bk|<1/2 \cr
                                                    0 & |\bk|\geq 1/2. \cr
                                                    \end{array} \right.  $$
To show that this function is infinitely differentiable it is enough to check 
that radial $k$-derivatives of all orders are zero at $k=1/2,$ both from 
the left and the right.  The above property implies that the inverse Fourier transform 
$$  G_0(\br) = \frac{1}{(2\pi)^d}\int d^dk \,\widehat{G}_0(\bk) \cos(\bk\bdot\br) $$
decays faster than any power $|\br|^{-p}$ as $|\br|\rightarrow\infty.$ Furthermore, 
$G_0(\br)$ is real-analytic, i.e. has a convergent power series representation 
with an infinite radius of convergence. The function $G_0(\br)$ is not everywhere 
positive, but $G(\br)=[G_0(\br)]^2$ is so. Also, $G(\br)$ is real-analytic and decays 
faster than any power as $|\br|\rightarrow\infty.$ It follows easily that 
$$ \widehat{G}(\bk) =  \frac{1}{(2\pi)^d}\int d^d p \,
                                    \widehat{G}_0(\bp) \widehat{G}_0(\bk-\bp)  $$  
is positive, $C^\infty$ and compactly supported in the ball $|\bk|<1.$ In Figs.\ref{fig:G} 
and \ref{fig:hatG} we have plotted the 1-dimensional versions 
of $G$ and $\hat{G}$ that follow from this construction, with the choice of 
$\widehat{G}_0$ given above.

As discussed in the text, it is often convenient to use filter kernels of  ``S-type'', which satisfy 
the condition (\ref{S-type}) in the text for some $\rho>1.$ To construct such a kernel, we 
first define a function $H(k)$ in Fourier space by 
$$                          H(k) = \left\{ \begin{array}{ll}
                                                    0 & 0\leq k\leq 1 \cr
                                                   \exp\left(-{{1}\over{(k-1)(\rho-k)}}\right) & 1<k<\rho \cr
                                                    0 & k \geq \rho \cr
                                                    \end{array} \right.  $$
which is $C^\infty$ and compactly supported in the interval $[1,\rho].$ We can now 
define $\widehat{G}(k)$ by
$$ \widehat{G}(k) = 1-\frac{1}{{\mathcal H}}\int_0^k dp\,H(p) $$
with ${\mathcal H}=\int_1^\rho dp\,H(p)>0.$ Clearly,
$$ \widehat{G}(k) = \left\{ \begin{array}{ll}
                                            1 & \mbox{if $k<1$} \cr
                                             0 & \mbox{if $k>\rho$} \cr
                                             \end{array}\right.  . $$
$\widehat{G}(\bk)=\widehat{G}(k)$ is also $C^\infty$ so that its $d$-dimensional 
inverse Fourier transform $G(r)$ decays faster than any power $r^{-p}$ as 
$r\rightarrow\infty.$ Furthermore,  $G(r)$ is real-analytic in $\br,$ because 
$\widehat{G}(k)$ is compactly supported. 

Note, however, that filter kernels $G$ of ``S-type'' cannot be non-negative, since
$$ \int d^dr\, |\br|^{2n}\,G(\br) = 
      \left. \left((-\bigtriangleup)^n \widehat{G}\right)(\bk)\right|_{\bk=0}=0 $$
for all positive integers $n.$  In that respect, they are similar to the sharp spectral
filter, whose physical-space kernel is proportional to a Bessel function that also takes 
on negative values. The primary advantage of the above construction is that the kernel $G$
is much better localized in physical space than is the kernel of the sharp-spectral 
filter, which decays only as a power-law $r^{-p}$ for $r\rightarrow\infty.$    

\newpage

\section{Band-Pass Decomposition of Quadratic Integrals}  

We show here how to construct a band-decomposition of quadratic integrals 
for a graded filter. Let us first make some definitions. We denote the filter 
function at length-scale $\ell_n=\rho^{-n}\ell$ by 
 $$ G_n(\br) = \ell_n^{-d} G(\br/\ell_n) $$
for a smooth filter kernel $G$ in space dimension $d.$ The corresponding low-pass filter is 
$$   \ol{f}_n(\bx) = (G_n*f)(\bx)=\int d\br \,G_n(\br) f(\bx+\br) $$ 
for any space function $f(\bx).$ In our development below we need also the 
{\it repeated low-pass filter}
$$  \ol{f}_{m,n}(\bx) = (G_n*G_{n+1}*\cdots *G_{m-1}*G_m*f)(\bx) $$
for any $n<m.$ Note that we have successively removed larger and larger length-scales,
so that this quantity is another type of low-pass filter at length-scale $\ell_n,$ for any $n<m.$
Finally, we define
$$ \tau_n(f,g)=\ol{(fg)_n}-\ol{f}_n\ol{g}_n. $$
This is the Germano (1992)  ``generalized central moment'' of 2nd-order, at length-scale $\ell_n$. 

We observe the following identity for arbitrary space functions $f,g$ and integer $N\geq 0$:
$$ \ol{(fg)}_{N,0}= \ol{f}_{N,0}\ol{g}_{N,0}+
      \sum_{n=1}^N \ol{(\tau_{n-1}(\ol{f}_{N,n},\ol{g}_{N,n}))}_{n-2,0} + \ol{(\tau_N(f,g))}_{N-1,0}. $$
It is not hard to see that successive terms cancel in this telescoping sum.  This identity
generalizes the well-known Germano identity \cite{Germano92}, to which it reduces when 
$N=1:$  $\tau_{1,0}(f,g)=\tau_0(\ol{f}_1,\ol{g}_1)+\ol{(\tau_1(f,g))}_0.$ 
Integration over space gives an exact band decomposition of a general quadratic integral,
\be \int fg = \int \ol{f}_{N,0}\ol{g}_{N,0}
   + \sum_{n=1}^N \int \tau_{n-1}(\ol{f}_{N,n},\ol{g}_{N,n}) + \int \tau_N(f,g).  \lb{band-d} \ee
Here we have used the fact that $\int \ol{f}=\int f$ for any filtered quantity. The physical interpretation 
of (\ref{band-d}) is that the first term corresponds to the contribution from length-scales $>\ell_0,$ the 
next terms for $n=1,...,N$ correspond, respectively, to contributions from length-scales between 
$\ell_{n-1}$ and $\ell_n,$ and the final term corresponds to contributions from length-scales $<\ell_N.$  

A simplified formula holds for filter kernels of  ``S-type''. In this case, $\ol{f}_{N,n}=\ol{f}_n$ 
for all $n<N$. Therefore,
$$  \int fg = \int \ol{f}_{0}\ol{g}_{0}
   + \sum_{n=1}^N \int \tau_{n-1}(\ol{f}_n,\ol{g}_n) + \int \tau_N(f,g)$$
This gives the formula (\ref{band-decomp}) of the text if one takes $f=g=u_i$ and then 
sums over $i.$

\bibliographystyle{unsrt}
\makeatletter
\renewcommand\@biblabel[1]{\textsuperscript{#1}}
\makeatother

\clearpage
\bibliography{turbulence}
\clearpage

\end{document}